\documentclass[twocolumn,epjc3]{svjour3}
\smartqed  % flush right qed marks, e.g. at end of proof
\RequirePackage[pdftex]{graphicx}
\RequirePackage{mathptmx}      % use Times fonts if available on your TeX system
\usepackage[utf8]{inputenc} % standard UTF-8 input encoding
\usepackage{amsmath,amssymb}
\usepackage[unicode,hypertexnames,setpagesize,%
    pdftex,%
    colorlinks,%
    citecolor=blue,%
    hyperindex,%
    plainpages=false,%
    bookmarksopen,%
    bookmarksnumbered%
  ]{hyperref}
\usepackage[numbers,square,comma,sort&compress]{natbib}

\usepackage{textcomp}

\usepackage{lineno,xcolor}
\usepackage{tikz}
\usetikzlibrary{arrows,shapes,positioning}

\usepackage{multirow}

\usepackage{bm}% bold math
\usepackage{xspace}
\journalname{Eur. Phys. J. C}
\begin{document}

\title{Testing Weak Equivalence Principle with Strongly Lensed Cosmic Transients}

\author{H. Yu \thanksref{addr1}
	\and
	F. Y. Wang\thanksref{e1,addr1,addr2}
}
\thankstext{e1}{Corresponding author: fayinwang@nju.edu.cn}
\institute{School of Astronomy and Space Science, Nanjing
	University, Nanjing 210093, China \label{addr1} \and Key Laboratory
	of Modern Astronomy and Astrophysics (Nanjing University), Ministry
	of Education, Nanjing 210093, China \label{addr2} }

\date{Received: date / Accepted: date}
\maketitle

\newcommand\jcap{{J. Cosmology Astropart. Phys.}}
\newcommand\na{{New A.}}
\def\nar{{\it New Astron. Rev. \,}}
\makeatletter
%
%  These Macros are taken from the AAS TeX macro package version 4.0.
%  Include this file in your LaTeX source only if you are not using
%  the AAS TeX macro package and need to resolve the macro definitions
%  in the BibTeX entries returned by the ADS abstract service.
%
%  If you plan not to use this file to resolve the journal macros
%  rather than the whole AAS TeX macro package, you should save the
%  file as ``aas_macros.sty'' and then include it in your paper by
%  using a construct such as:
%	\documentstyle[11pt,aas_macros]{article}
%
%  For more information on the AASTeX macro package, please see the URL
%	http://www.aas.org/publications/aastex.html
%  For more information about ADS abstract server, please see the URL
%	http://adswww.harvard.edu/ads_abstracts.html
%

% Abbreviations for journals.  The object here is to provide authors
% with convenient shorthands for the most "popular" (often-cited)
% journals; the author can use these markup tags without being concerned
% about the exact form of the journal abbreviation, or its formatting.
% It is up to the keeper of the macros to make sure the macros expand
% to the proper text.  If macro package writers agree to all use the
% same TeX command name, authors only have to remember one thing, and
% the style file will take care of editorial preferences.  This also
% applies when a single journal decides to revamp its abbreviating
% scheme, as happened with the ApJ (Abt 1991).

\let\jnl@style=\rm
\def\ref@jnl#1{{\jnl@style#1}}

\def\aj{\ref@jnl{AJ}}                   % Astronomical Journal
\def\araa{\ref@jnl{ARA\&A}}             % Annual Review of Astron and Astrophys
\def\apj{\ref@jnl{ApJ}}                 % Astrophysical Journal
\def\apjl{\ref@jnl{ApJ}}                % Astrophysical Journal, Letters
\def\apjs{\ref@jnl{ApJS}}               % Astrophysical Journal, Supplement
\def\ao{\ref@jnl{Appl.~Opt.}}           % Applied Optics
\def\apss{\ref@jnl{Ap\&SS}}             % Astrophysics and Space Science
\def\aap{\ref@jnl{A\&A}}                % Astronomy and Astrophysics
\def\aapr{\ref@jnl{A\&A~Rev.}}          % Astronomy and Astrophysics Reviews
\def\aaps{\ref@jnl{A\&AS}}              % Astronomy and Astrophysics, Supplement
\def\azh{\ref@jnl{AZh}}                 % Astronomicheskii Zhurnal
\def\baas{\ref@jnl{BAAS}}               % Bulletin of the AAS
\def\jrasc{\ref@jnl{JRASC}}             % Journal of the RAS of Canada
\def\memras{\ref@jnl{MmRAS}}            % Memoirs of the RAS
\def\mnras{\ref@jnl{MNRAS}}             % Monthly Notices of the RAS
\def\pra{\ref@jnl{Phys.~Rev.~A}}        % Physical Review A: General Physics
\def\prb{\ref@jnl{Phys.~Rev.~B}}        % Physical Review B: Solid State
\def\prc{\ref@jnl{Phys.~Rev.~C}}        % Physical Review C
\def\prd{\ref@jnl{Phys.~Rev.~D}}        % Physical Review D
\def\pre{\ref@jnl{Phys.~Rev.~E}}        % Physical Review E
\def\prl{\ref@jnl{Phys.~Rev.~Lett.}}    % Physical Review Letters
\def\pasp{\ref@jnl{PASP}}               % Publications of the ASP
\def\pasj{\ref@jnl{PASJ}}               % Publications of the ASJ
\def\qjras{\ref@jnl{QJRAS}}             % Quarterly Journal of the RAS
\def\skytel{\ref@jnl{S\&T}}             % Sky and Telescope
\def\solphys{\ref@jnl{Sol.~Phys.}}      % Solar Physics
\def\sovast{\ref@jnl{Soviet~Ast.}}      % Soviet Astronomy
\def\ssr{\ref@jnl{Space~Sci.~Rev.}}     % Space Science Reviews
\def\zap{\ref@jnl{ZAp}}                 % Zeitschrift fuer Astrophysik
\def\nat{\ref@jnl{Nature}}              % Nature
\def\iaucirc{\ref@jnl{IAU~Circ.}}       % IAU Cirulars
\def\aplett{\ref@jnl{Astrophys.~Lett.}} % Astrophysics Letters
\def\apspr{\ref@jnl{Astrophys.~Space~Phys.~Res.}}
                % Astrophysics Space Physics Research
\def\bain{\ref@jnl{Bull.~Astron.~Inst.~Netherlands}} 
                % Bulletin Astronomical Institute of the Netherlands
\def\fcp{\ref@jnl{Fund.~Cosmic~Phys.}}  % Fundamental Cosmic Physics
\def\gca{\ref@jnl{Geochim.~Cosmochim.~Acta}}   % Geochimica Cosmochimica Acta
\def\grl{\ref@jnl{Geophys.~Res.~Lett.}} % Geophysics Research Letters
\def\jcp{\ref@jnl{J.~Chem.~Phys.}}      % Journal of Chemical Physics
\def\jgr{\ref@jnl{J.~Geophys.~Res.}}    % Journal of Geophysics Research
\def\jqsrt{\ref@jnl{J.~Quant.~Spec.~Radiat.~Transf.}}
                % Journal of Quantitiative Spectroscopy and Radiative Transfer
\def\memsai{\ref@jnl{Mem.~Soc.~Astron.~Italiana}}
                % Mem. Societa Astronomica Italiana
\def\nphysa{\ref@jnl{Nucl.~Phys.~A}}   % Nuclear Physics A
\def\physrep{\ref@jnl{Phys.~Rep.}}   % Physics Reports
\def\physscr{\ref@jnl{Phys.~Scr}}   % Physica Scripta
\def\planss{\ref@jnl{Planet.~Space~Sci.}}   % Planetary Space Science
\def\procspie{\ref@jnl{Proc.~SPIE}}   % Proceedings of the SPIE

\let\astap=\aap
\let\apjlett=\apjl
\let\apjsupp=\apjs
\let\applopt=\ao

\makeatother

\setcounter{tocdepth}{4}
\maketitle

\begin{abstract}
	{Current constraints on Einstein's weak equivalence principle (WEP) utilize the observed time delay between correlated particles of
		astronomical sources. However, the intrinsic time delay due
			to particle emission time and the time delays caused by potential
			Lorentz invariance violation and non-zero rest mass of photons are
			simply omitted in previous studies. Here we propose a robust method
			to test WEP using strongly lensed cosmic transients, which can
			naturally overcome these time delays. This can be achieved by
		comparing the time delays between lensed images seen in different
		energy bands or in gravitational waves (GWs) and their
		electromagnetic (EM) counterparts. The power of our method mainly
		depends on the timing accuracy of cosmic transient and the strong
		lensing time delay. If the time delay of cosmic transient can be
		measured with accuracy about 0.1 s (e.g. gamma-ray bursts), we show
		that the upper limit on the differences of the parameterized
		post-Newtonian parameter $\gamma$ value is $\Delta \gamma<10^{-7}$
		with a one-month strong lensing time delay event. This accuracy of
		WEP can be improved by several orders, if the lens is galaxy cluster
		and the strongly lensed cosmic transients have much shorter
		duration, such as fast radio bursts.}
\end{abstract}
\maketitle

\section{Introduction}\label{sec:intro}
The Einstein's weak equivalence principle (WEP) is one of the
cornerstone of general relativity. It states that the trajectories
of any freely falling, uncharged test bodies are independent of
their energy, internal structure, or composition
\cite{Will2014LRR....17....4W}. In the theory of parameterized post-Newtonian
(PPN), the WEP requires that the PPN parameter $\gamma$ of different
particles or the same type of particle with different energies
(hereafter, ``different particles'' represents both of cases) should
be the same. WEP has been tested by time delay of different
particles. In 1964, \cite{Shapiro1964PhRvL..13..789S} proposed that
one could use the time delays between transmission of radar pulses
towards either of the inner planets and detection of the echoes. The
Shapiro time delay can be formulated as $t_{\rm
	Shapiro}=-\frac{1+\gamma}{c^3}\int_{r_e}^{r_0}\Psi(r){\rm d}r$,
where $r_e$ and $r_0$ are the emitting position of the particle and
the position of observer respectively, $c$ is the speed of light,
and $\Psi(r)$ is the gravitational potential. If the WEP is
violated, the arrive times of two different particles emitted
simultaneously travel in a same gravitational potential will be
different. The relative Shapiro time delay is given by
\begin{equation}\label{ShapiroDelay}
\Delta t_{\rm gra} = \frac{|\gamma_1-\gamma_2|}{c^3}\int_{r_e}^{r_0}\Psi(r){\rm
	d}r.
\end{equation}

Up to date, the observed time delays of different type
particles (e.g. photons, neutrinos, or gravitational waves), or the
same type of particles with different energies have been used to
test WEP, such as the different arrival times of photons and
neutrinos from SN1987A
\cite{Krauss1988PhRvL..60..176K,Longo1988PhRvL..60..173L}, the time
delay of a PeV-energy neutrino event associated with a giant flare
of the blazar PKS B1424-418 \cite{Wang2016PhRvL.116o1101W}, the
photons in different energy bands of gamma-ray bursts (GRBs)
\cite{Gao2015ApJ...810..121G,2017MNRAS.469L..36Y}, polarized photons of GRBs
\cite{2017MNRAS.469L..36Y}, radio signals at different frequency
bands of fast radio bursts (FRBs)
\cite{Wei2015PhRvL.115z1101W,Nusser2016ApJ...821L...2N,Zhang2016arXiv160104558Z}
and the Crab pulsar \cite{Yang2016PhRvD..94j1501Y}, and
gravitational wave (GW) sources
\cite{Wu2016PhRvD..94b4061W,Kahya2016PhLB..756..265K,Abbott2017ApJ...848L..13A,Wei2017JCAP...11..035W}. However,
the observed time delay should include several terms:
\begin{equation}
\Delta t_{\rm obs}=\Delta t_{\rm int}+\Delta t_{\rm LIV}+\Delta
t_{\rm spec}+\Delta t_{\rm DM}+\Delta t_{\rm gra},
\end{equation}
where $\Delta t_{\rm int}$ is the intrinsic time delay between two
particles due to different emission times, $\Delta t_{\rm LIV}$ is
the time delay caused by Lorentz invariance violation, $\Delta
t_{\rm spec}$ is the potential time delay due to photons have a rest
mass, and $\Delta t_{\rm DM}$ is the time delay contributed by the
dispersion of the line-of-sight free electrons.

In the high-energy range (from keV to GeV), $\Delta t_{\rm
		DM}$ can be negligible. However, most of previous works omitted all
	of other potential effects based on following assumptions. First,
	the observed time delay is mainly attributed to the gravitational
	potential so the time delays caused by other effects are omitted.
	Second, the time delays caused by other effects and $\Delta t_{\rm
		gra}$ have the same sign. So they can give an upper limit of the
	violation of WEP. These two assumptions are unreasonable. For the
	first assumption, in some cases, the $\Delta t_{\rm int}$ term would
	dominate the observed time delay especially when the observed time
	delay is very small, which leads to a non-physical constraint on the
	WEP. In addition, $\Delta t_{\rm spec}$ and $\Delta t_{\rm LIV}$ are
	strongly correlated with $\Delta t_{\rm gra}$ in the total observed
	time delay. For the second assumption, in a special case that the
	signs of $\Delta t_{\rm other}$ caused by other effects and $\Delta
	t_{\rm gra}$ are opposite (e.g. $\Delta t_{\rm other}=-0.99$ s and
	$\Delta t_{\rm gra}=1$ s), a much smaller time delay is obtained.
	This will lead to a much tighter, but incorrect constraint on WEP.
	Therefore, the intrinsic time delay $\Delta t_{\rm int}$ and $\Delta t_{\rm other}$ can
	severely limit the ability of this method. To avoid these effects, we propose that the strongly
	lensed cosmic transients can be used to test WEP. This is the first
	time to correct the other effects in WEP constraint. The time delay
	due to strong gravitational lensing between different particles is a
	powerful tool, which has been used to test the Lorentz-invariance
	violation \cite{Biesiada2009MNRAS.396..946B} and the speed of GW
	\cite{Collett2017PhysRevLett.118.091101,Fan2017PhysRevLett.118.091102}.

\section{The Method and Constraints on WEP}
Gravitational lens is a prediction of the general relativity. After
the first observation example of gravitational lensing, the quasar
QSO 0957+561A, B \cite{Walsh1979Natur.279..381W}, it has become a
powerful tool in many fields of astronomy, such as probing dark
matter halo, large scale structures, Hubble constant, and parameters
of universe. Generally, there will be multiple images of the source
when it is strongly lensed. The differences of arrival times for
images are caused by the Shapiro time delay and geometric delay due
to the bending of light rays \cite{Narayan1996astro.ph..6001N}. In a
general model of the lens, the time delay of the images relative to
the case that source, lens and image are on a straight line is
\begin{equation}\label{eq:LensTimeDelay}
t(\bm{\theta})=\frac{1+z_l}{c}\frac{d_ld_s}{d_{ls}}[\frac{1}{2}(\bm{\theta}-\bm{\beta})^2-\psi(\bm{\theta})],
\end{equation}
where $\bm{\theta}$ and $\bm{\beta}$ are the position vectors of the
image and source, $z_l$ is the redshift of the lens, $d_l$ and $d_s$
are the angular diameter distances of the lens and source, $d_{ls}$
is the angular diameter distance from the lens to source and the
$\psi(\bm{\theta})=\frac{d_{ls}}{d_ld_s}\frac{1+\gamma}{c^2}\int\Psi(d_l\bm{\theta},z)dz$
is the projected gravitational potential \cite{Narayan1996astro.ph..6001N}.
Actually, $t(\bm{\theta})$ can be divided into $t_{\rm
	geo}=\frac{1+z_l}{c}\frac{d_ld_s}{d_{ls}}\frac{1}{2}(\bm{\theta}-\bm{\beta})^2$
and $t_{\rm
	gra}=\frac{1+z_l}{c}\frac{d_ld_s}{d_{ls}}\psi(\bm{\theta})$, which
are the geometric time delay and Shapiro time delay, respectively.

\begin{figure}
	%  \centering
	\includegraphics[width=0.5\textwidth]{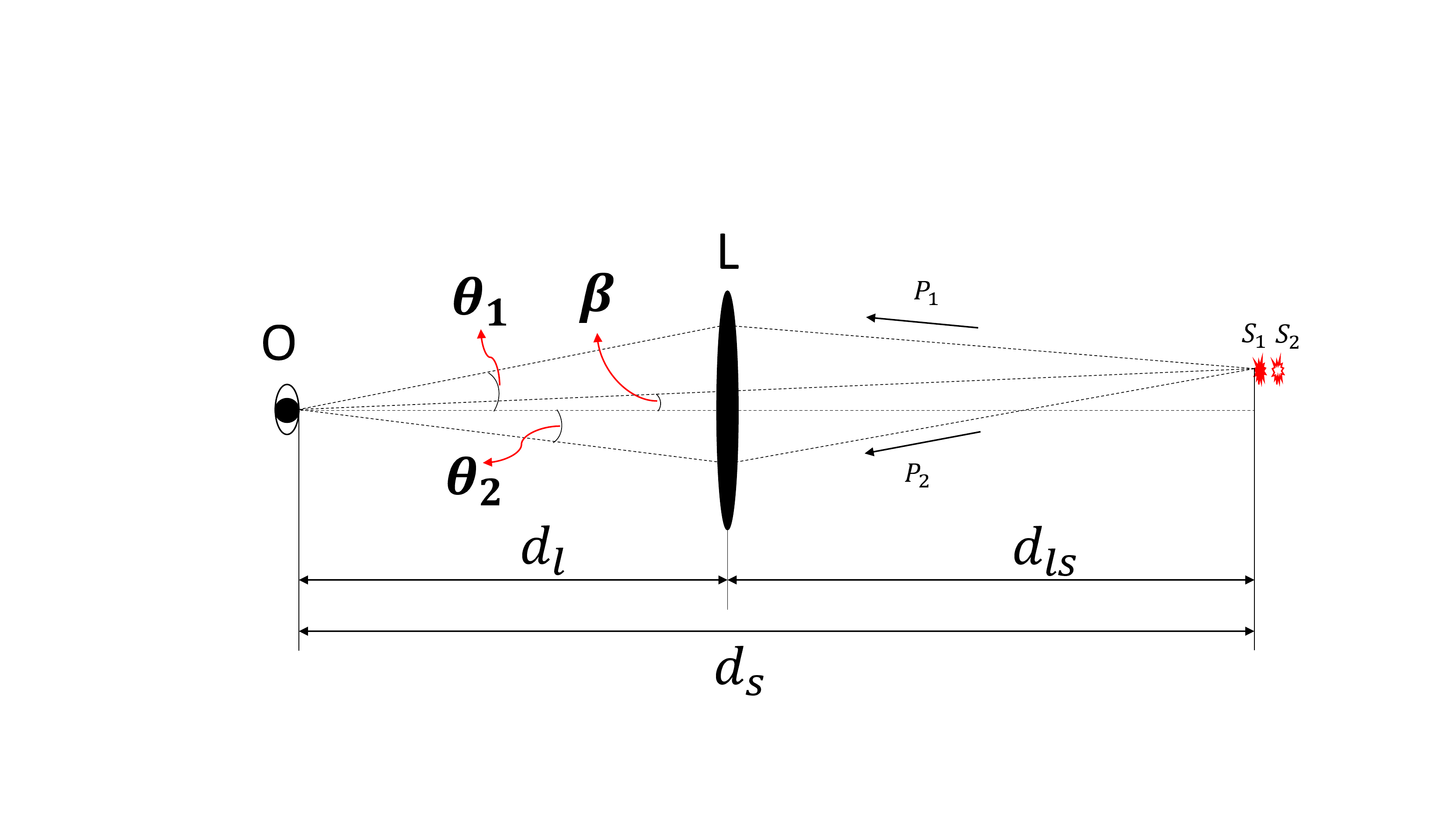}\\
	\caption{The geometry of gravitational lens considered here. Illustration of a strongly lensed cosmic transient which can be used to test WEP.}\label{figLens}
\end{figure}

We show a strong lens of a bright cosmic transient in Figure
\ref{figLens}. In this figure, $O$ and $L$ are the observer and lens
object. $S_1$ and $S_2$ are the two signals which are associated
with each other in this transient event, for example a GW and its
electromagnetic (EM) counterpart or particles at two different
energy bands in a cosmic explosion. $d_l$, $d_s$, $d_{ls}$ and
$\bm{\beta}$ are the same as those in equation
(\ref{eq:LensTimeDelay}). $\bm{\theta_1}$ and $\bm{\theta_2}$ are
the positions of two images formed by gravitational lens. $P_1$ and
$P_2$ are the trajectories of the light rays of the two images. We
assume that the intrinsic time delay between $S_1$ and $S_2$ is
$\Delta t_{\rm int}$ ($S_1$ is earlier than $S_2$) and the first and
second signals of $S_1$ arrive at $t_{11}$ and $t_{12}$,
respectively. Here we just assume that the $\Delta t_{\rm
		spec}$ and $\Delta t_{\rm LIV}$ can be omitted and discuss them in
	the following section. If the WEP is valid, one can expect that the
first and second signals of $S_2$ will arrive at
$t_{21}=t_{11}+(1+z_s)\Delta t_{\rm int}$ and
$t_{22}=t_{12}+(1+z_s)\Delta t_{\rm int}$, where $z_s$ is the
redshift of the source. If there is some small violation of WEP,
positions of the images have small changes $\delta\theta_{11}$,
$\delta\theta_{12}$, $\delta\theta_{21}$, $\delta\theta_{22}$ and
the arrival time of the signals become into $t_{11}^\prime$,
$t_{12}^\prime$, $t_{21}^\prime$, and $t_{22}^\prime$. Similar as
\cite{Baker2017PhRvD..95f3512B}, we perform the Taylor expansion of
these new arrival times and consider the first order term
\begin{eqnarray*}
	t_{11}^\prime &=& t_{11}+\frac{\partial t}{\partial\theta}|_{\theta_1}\delta\theta_{11}+\frac{\partial t}{\partial\gamma}|_{\theta_1,\gamma_0}(\gamma_1-\gamma_0), \\
	t_{12}^\prime &=& t_{12}+\frac{\partial t}{\partial\theta}|_{\theta_2}\delta\theta_{12}+\frac{\partial t}{\partial\gamma}|_{\theta_2,\gamma_0}(\gamma_1-\gamma_0), \\
	t_{21}^\prime &=& t_{21}+\frac{\partial t}{\partial\theta}|_{\theta_1}\delta\theta_{21}+\frac{\partial t}{\partial\gamma}|_{\theta_1,\gamma_0}(\gamma_2-\gamma_0), \\
	t_{22}^\prime &=& t_{22}+\frac{\partial t}{\partial\theta}|_{\theta_2}\delta\theta_{22}+\frac{\partial t}{\partial\gamma}|_{\theta_2,\gamma_0}(\gamma_2-\gamma_0).
\end{eqnarray*}
Because the Fermat's principle, which we assume is still valid in this case, requires that the lensed image
position makes the travel time stationary, the $\frac{\partial
	t}{\partial\theta}|_{\theta_1}$ and $\frac{\partial
	t}{\partial\theta}|_{\theta_2}$ are equal to $0$ and all the second
terms of the right parts of these equations must be vanished. Then comparing $t_{21}^\prime-t_{11}^\prime$
and $t_{22}^\prime-t_{12}^\prime$, the effect of the intrinsic time delay $t_{\rm
	int}$ can be removed naturally. The difference between $\gamma_1$
and $\gamma_2$ is
\begin{equation}\label{eq:deltaGamma}
\Delta \gamma\equiv|\gamma_1-\gamma_2|\leq 2(1+\alpha)\frac{|(t_{22}^\prime-t_{12}^\prime)-(t_{21}^\prime-t_{11}^\prime)|}{|t_{22}^\prime-t_{21}^\prime|},
\end{equation}
where $\alpha=\Delta t_{\rm geo}/\Delta t_{\rm gra}$ is the ratio of
time delays caused by geometric effect and Shapiro delay effect.
Hereafter, we define `time delay' as $t_{22}^\prime-t_{12}^\prime$ or
$t_{21}^\prime-t_{11}^\prime$, which is the time difference between different
particles in the same light path. The `strong lensing time delay' is
defined as $t_{22}^\prime-t_{21}^\prime$ or $t_{12}^\prime-t_{11}^\prime$, which is the time
delay between two different paths due to strong lensing. The value
of $\alpha$ depends on the choice of lens model.

For the purpose of illustrating our method, we use the singular
isothermal sphere (SIS) model, which has been proved to be a
reliable model for lenses. In the SIS model, the distribution of
stars and other mass components in galaxies are thought to be like
that of particles in idea gas. The projected potential and Einstein
radius of a SIS lens are
$\psi(\xi)=\frac{d_{ls}}{d_s}\frac{4\pi\sigma_v^2}{c^2}|\xi|$ and
$\theta_E=4\pi\frac{\sigma_v^2}{c^2}\frac{d_{ls}}{d_s}$
respectively, where $\sigma_v$ is the velocity dispersion and $\xi$
is the angular distance from the center of the lens
\cite{Narayan1996astro.ph..6001N}. If the lensing is strong,
$\beta<\theta_E$, there are two images of the source at the
positions $\theta_{\pm}=\beta\pm\theta_E$. Then, from equation
(\ref{eq:LensTimeDelay}), the time delay between the two images is
\begin{equation}\label{eq:deltaT_SIS}
\Delta T_{\rm SIS} = 2\beta\theta_E\frac{1+z_l}{c}\frac{d_ld_s}{d_{ls}},
\end{equation}
which is all caused by the Shapiro time delay effect (i.e. $\Delta
T_{\rm SIS}=t_{\rm gra,\,SIS}$). The difference in arrival time of
two images caused by geometric time delay is $\Delta t_{\rm geo}=0$
which means that the length of the different light trajectories of
the two images $P_1$ and $P_2$ are equal to each other. It
should be pointed out that the SIS model is an idealized lens model.
The real strong lensing by a galaxy is likely not to be a SIS case.
For a certain strong lensing event, if there are other observation
of the properties of the lens object, we can calculate the value of
$\alpha$ in a suitable lens model. Generally, both of $\Delta t_{\rm
	geo}$ and $\Delta t_{\rm gra}$ are on the order of $GM/c^3$, where
$M$ is the mass of lens \cite{Weinberg2008cosm.book.....W}.
Therefore, one can expect that the parameter $\alpha$ should be on
the order of one.

GRBs are promising tools to constrain the WEP for photons
	with different energies \cite{Gao2015ApJ...810..121G}. Due to their
	large luminosities, GRBs can be observed at very high redshifts
	\cite{Wang2015NewAR..67....1W}. Therefore, there is much more
	possibility for a GRB to be lensed by a galaxy or galaxy cluster in
	the universe. Due to the success of BATSE, Swift and Fermi
	satellites, the number of detected GRBs keeps increasing. There
	are lots of works to search the potential strong
	lensed GRBs in several GRB's catalogs. Although no such event was
	ever found \cite{Veres2009arXiv0912.3928V,Li2014SCPMA..57.1592L}. Li
	\& Li (2014) searched the potential lensing events in BATSE GRB data. They
	found four candidates.
	The second couple of GRBs, 2044 and 2368, has closest properties of
	a strong lensed GRB event. We use this GRB as an example, although
	they excluded the possibility in their work
	\cite{Li2014SCPMA..57.1592L}. The flux ratios in four considered
	energy channels seem similar. In addition, the angular separation of
	them is $\Delta\theta=3.88^\circ$ while the location uncertainties
	of them are $2.88^\circ$ and $6.06^\circ$. The detected time delay
	between 2044 and 2368 is about $1.77\times10^7$ s and the time
	delays of the photons in energy channels of 25-60 keV and 60-110 keV
	are $0.085\pm0.042$ s and $1.730\pm0.162$ s for 2044 and 2368
	respectively. Assuming there is a lensed GRB event with similar time
	delay, we constrain the WEP with our method. From equation
	(\ref{eq:deltaGamma}), the constraint on the violation of WEP
	between the photons in these two energy bands is
	$\Delta\gamma\leq2(1+\alpha)\frac{1.730-0.085}{1.77\times10^7}=1.86(1+\alpha)\times10^{-7}$.
	If we choose the SIS lens model, $\alpha=0$, it has
	$\Delta\gamma\leq1.86\times10^{-7}$. For Fermi GBM, the expected
	time to observe one lensed GRB is about 11 years
	\cite{Li2014SCPMA..57.1592L}. Since Fermi GBM has served about 9
	years and will be in service for another more than 10 years, it is
	reasonable to expect a lensed GRB event in the operating period of
	Fermi GBM.

\section{Discussion}
In this section, we discuss two potential biases. The first
	point is the time delay potentially caused by the non-zero rest mass
	of photons $\Delta t_{\rm spec}$ and the LIV $\Delta t_{\rm LIV}$.
	Generally, these two effects are similar since both of them will
	cause the energy-dependent speed of photons. If the photon has
	non-zero rest mass, then the higher-energy photons will travel
	faster than lower-energy photons. On the contrary, LIV will lead to
	an opposite effect that higher energy photons have slower traveling
	speed because the so-called vacuum dispersion
	\cite{Amelino-Camelia1997IJMPA..12..607A,Ellis2008PhLB..665..412E,Ellis2011IJMPA..26.2243E,Kostelecky2008ApJ...689L...1K}.
	Therefore, the time delay terms $\Delta t_{\rm spec}$ and $\Delta
	t_{\rm LIV}$ are both caused by the potential difference of
	traveling speed of photons with different energies.

In figure \ref{figLens}, there are two traveling paths and
	for each path there are two images formed by two different kinds of
	particles. Let's assume the speeds of those two different particles
	are $v_1$ and $v_2$ and the lengths of the two traveling paths are
	$L_1$ and $L_2$, respectively. Therefore, the time delays caused by
	this effect are $L_1/v_1-L_1/v_2$ and $L_2/v_1-L_2/v_2$
	respectively. Their contribution to the difference of time delays
	$|(t_{22}^\prime-t_{12}^\prime)-(t_{21}^\prime-t_{11}^\prime)|$ which we use in equation
	(\ref{eq:deltaGamma}) to constrain the WEP is
	$\frac{(L_1-L_2)(v_2-v_1)}{v_1v_2}$. Considering a strong lensing
	system with strong lensing time delay is about 1 year, and the time
	delays of different particles $t_{22}^\prime-t_{12}^\prime$ and $t_{21}^\prime-t_{11}^\prime$
	are about 1 second, the difference of time delays
	$|(t_{22}^\prime-t_{12}^\prime)-(t_{21}^\prime-t_{11}^\prime)|$ should be also at order of 1
	second. For this kind of strong lensing system, the difference
	between $L_1$ and $L_2$ should be at a order of 1 light year.
	However the distance of a typical cosmic source should be at a order
	of 1 Gpc, so it has
\begin{equation}
\frac{(L_1-L_2)(v_2-v_1)}{v_1v_2}/(\frac{L_1}{v_1}-\frac{L_1}{v_2}) = \frac{L_1-L_2}{L_1} \sim \frac{1 {\rm ly}}{1 {\rm Gpc}}\sim 10^{-9}.
\end{equation}
Therefore, even though the time delay caused by non-zero
	rest mass of photons and LIV effect is hundreds of times larger than
	the observed time delay between two different particles,
	$\frac{(L_1-L_2)(v_2-v_1)}{v_1v_2}$ only contributes very small
	part of  $|(t_{22}^\prime-t_{12}^\prime)-(t_{21}^\prime-t_{11}^\prime)|$. Therefore, it is
	reasonable to omit the terms $\Delta t_{\rm spec}$ and $\Delta
	t_{\rm LIV}$, which means our method can also exclude the potential
	effects of non-zero rest mass of photons and LIV.

The second point is that the two traveling paths will
	perhaps introduce some other contributions to the observed time
	delay. Actually, for a typical strong lensing system, the Einstein
	radius is about $4\times10^{-6}
	(\frac{M}{10^{11}M_\odot})^{0.5}(\frac{D}{1\rm
		Gpc})^{-0.5}$\cite{Narayan1996astro.ph..6001N}, which is very
	smaller compared to the relativistic beaming angle
	$\theta\sim\Gamma^{-1}\sim10^{-3}$ where $\Gamma$ is the Lorentz
	factor GRB's jet \cite{Granot2002ApJ...570L..61G}. Therefore, the
	difference of the special-relativistic boost factor caused by
	different viewing angles can be omitted reasonably. In addition, the
	different paths may also lead to different effects of gravitational potential along the paths, such as the weak lensing by the large scale structure and also the effects of our Galaxy and local galaxy cluster, which may also contribute into the observed time
	delay. However, from equation (\ref{ShapiroDelay}), the difference between the Shapiro time delays for two paths relies on the total difference of the gravity potential along the whole paths which includes the gravitational potential of large scale structure, the local galaxy cluster and also our galaxy. Therefore, the effect is considered even though there is some small difference of gravitational potential along two different paths.

\section{Conclusions}
In this paper, we have proposed a method to constrain the violation
of WEP with strongly lensed cosmic transients. Our method dose not
need to make any assumption and understand physical mechanism of the
transient. Moreover, because it utilizes the difference of time
delays of the two lensing images, the potential effect of intrinsic
time delay $\Delta t_{\rm int}$, LIV time delay $\Delta t_{\rm LIV}$
and non-zero rest mass time delay $\Delta t_{\rm spec}$ can be
naturally removed. By analyzing the properties of time delay of
gravitational lens, we find that the parameter $\alpha$, which
represents the ratio of time delays between two lensing images
caused by geometric and Shapiro delay effects, will be zero for the
SIS lens model. Therefore, the lengths of light ray paths for two
lensing images are the same in the SIS lens model, which means any
test depending on the difference of the path length can not work in
SIS lens model, such as testing the difference between the speeds of
light and GW. Even though for a realistic lens, one should consider
the parameter $\alpha$ when constraining the speed of GW. Otherwise
an unreasonable tighter constraint will be obtained.

Due to the number of GRBs increasing, the detection of a
	strong lensed GRB is promising. Although there is still no such
	event at present. Assuming a strong lensed GRB with time delay
	between images is about $1.77\times10^7$ s and the time delays of
	the photons in two energy channels are $0.085\pm0.042$ s and
	$1.730\pm0.162$ s, one can give a constraint on WEP at about
	$10^{-7}$ level. Besides GRBs, the lensed FRBs and the GW events
	with their EM counterparts are also potential candidates to test the
	WEP using our method
	\cite{Abbott2016PhRvL.116x1103A,2017Natur.541...58C,2017ApJ...834L...7T}. Interestingly, lensing of FRBs
	has been proposed to probe dark matter \cite{Munoz2016PhRvL.117i1301M,Wang2018arXiv180107360W}.

From equation (\ref{eq:deltaGamma}), with the time delay
measurements of a single strong gravitational lensing event one can
give a tight constrain on the $\Delta \gamma$. It can also be found
that the efficiency of the constraint on WEP is proportional to the
accuracy of time delay measurement and inversely proportional to the
strong lensing time delay. A typical strong lensing by a galaxy will
give multiple images with days to months strong lensing time delays
\cite{Oguri2010MNRAS.405.2579O}. Because the strong lensing time
delay is proportional to the mass of lens, the time delay due to
galaxy cluster lens is much longer \cite{Kelly2015Sci...347.1123K},
which will give much more strict constraint on $\Delta \gamma$. If a
several-months strong lensing time delay is observed and the time
delay of two images is about 0.1 s, the constraint on the violation
of WEP will be up to $\Delta\gamma<10^{-8}$. For the measurement of
time delay between two different particles, it depends on the
accuracy of timing measurement of the transient event. The
accuracies of time measurement of FRB, GW and GRB are about 0.01 ms
\cite{Champion2016MNRAS.460L..30C}, $10^{-4}$ ms
\cite{Abbott2016PhRvL.116x1103A} and 0.1 s
\cite{Abdo2010ApJS..187..460A}. Recently, the detections of
	GW170817 and its electromagnetic counterparts
	\cite{Abbott2017ApJ...848L..13A,Abbott2017ApJ...848L..12A} also
	encourage us to make a more reliable constraint on the WEP using
	strong lensing of GWs and electromagnetic counterparts. Therefore,
the accuracy of WEP can be improved by several orders of magnitude
in the future, if the lens is galaxy cluster and the strongly lensed
cosmic transients have much more precise time delay measurement,
such as the GW events and FRBs.

%\tableofcontents

%\input{}
%Text with citations \cite{RefB} and \cite{RefJ}.

%as required. Don't forget to give each section
%and subsection a unique label (see Sect.~\ref{sec:1}).
%\paragraph{Paragraph headings} Use paragraph headings as needed.
%\begin{equation}
%\end{equation}

%%%%%%%%%%%%%%%%%%%%%%%%%%%%%%%%%%%%%%%%%%%%%%%%%%%%%%%%
%% For one-column wide figures use
%\begin{figure}
%% Use the relevant command to insert your figure file.
%% For example, with the graphicx package use
%  \includegraphics{example.eps}
%\end{figure}
%
%%%%%%%%%%%%%%%%%%%%%%%%%%%%%%%%%%%%%%%%%%%%%%%%%%%%%%%%
%% For two-column wide figures use
%\begin{figure*}
%% Use the relevant command to insert your figure file.
%% For example, with the graphicx package use
%  \includegraphics[width=0.75\textwidth]{example.eps}
%\end{figure*}
%\begin{table}
%\begin{tabular}{lll}
%\hline\noalign{\smallskip}
%\end{tabular}
%\end{table}

%%%%%%%%%%%%%%%%%%%%%%%%%%%%%%%%%%%%%%%%%%%%%%%%%%%%%%%%
\begin{acknowledgements}
We thank the anonymous referee for detailed comments and
suggestions. This work is supported by the National Basic Research
Program of China (973 Program, grant No. 2014CB845800), the National
Natural Science Foundation of China (grant Nos. 11422325, and
11373022), and the Excellent Youth Foundation of Jiangsu Province
(BK20140016).
\end{acknowledgements}


%\bibliography{bibfile.bib}

\end{document}